# An Inertial Cell Model for the Drag Force in Multiphase Flow

G. B. Tupper[1,2,3], I. Govender[1,2], and A.N. Mainza[2]

[1]*Applied Physics Group, Department of physics, University of Cape Town*

[2]*Centre for Minerals Research, Department of Chemical Engineering, University of Cape Town*

[3]*Associate Member, National Institute for Theoretical Physics*

*Email contact of corresponding author*: gary.tupper@uct.ac.za

ABSTRACT: A new model for the drag coefficient of a sphere in a concentrated system is described. It is based upon a cell-averaged model for the Stokes regime, combined with a physically motivated extrapolation to arbitrary Reynolds number. It can be used as an alternative to the isolated particle drag coefficient in Euler–Lagrange modelling of solid-liquid multiphase flow. The corresponding drag force also provides a dynamic bed equation for use in Euler- Euler modelling.

## 0. Introduction

Multiphase flows, in which solids comprise at least one dispersed phase, are central to many problems in chemical engineering, ranging from fluidized bed reactors to the cyclones and mills employed in comminution. Computational Fluid Mechanics (CFD) is a ubiquitous tool for the study of such flows; for a review see van Wachem and Almstedt (2003). Now, a central issue in multiphase modelling is the interphase momentum transfer, or drag model, need for closure. Taking the case of spheres in an incompressible Newtonian fluid the key question can be paraphrased as: what is the dependence of the dimensionless drag coefficient $C_D(R_e, \varepsilon)$ on both the local particle Reynolds number $R_e$ and also on the local fluid volume fraction or voidage $\varepsilon$ ? What is wanted is a drag coefficient that is a continuous function of the voidage and has the isolated sphere as the limiting case, i.e. $C_D(R_e, \varepsilon \to 1) \to C_D(R_e)$.

If one is dealing with say a fluidized bed problem, and adopts an Eulerian-Eulerian approach to solid-fluid multiphase flow in which the particle phase is modelled as a continuum, the question settles down to a choice between the various standard drag models that are part of commercial as well as open source CFD packages. The Wen and Yu (1966), Syamlal and O'Brien (1987), and the Di Felice (1994) models all make use of the classic Richardson-Zaki (1954) study of settling velocities. The Wen-Yu model equation follows the Richardson-Zaki (1954) prescription of modifying the drag on an isolated sphere $C_D(R_e)$ by a factor of the relative settling velocity $v_r = \varepsilon^{-\alpha}$ to account for the influence of neighbouring particles on the flow around it. Wen and Yu take the index $\alpha = 2.65$ as constant whereas in Di Felice $\alpha = \alpha(R_e)$ is fitted. Syamlal and O'Brien build in settling data more intimately via $C_D(R_e/v_r)/v_r^2$. Irrespective, the fact that settling and fluidization are just the same process viewed from different frames, Richardson-Zaki (1954), means that using one of these gives reasonable confidence in the drag closure aspect of the fluidized bed model.

If one instead adopts the Eulerian – Lagrangian approach one is generally left with a choice between an isolated particle drag model and that of Gidaspow (1994). The use of the isolated particle drag model is appropriate in the traditional application of Eulerian – Lagrangian modelling to small solids concentration where collisions are relatively unimportant and collective effects should arise out of the simulation rather than being built into the particle-fluid coupling via the drag force. As reviewed by Deen, Annaland, Van der Hoef, and Kuipers, (2007), the steady advance of computational power means that it is also possible to pursue coupled Discrete Element Method (DEM)-CFD into regimes where the solid volume fraction is large. In so far as the particle Reynolds number is sufficiently high the use of an isolated particle drag model may still be justified, an example being the Chu, Wang, Yu and Vince (2009) simulation of dense medium cyclones.

On the other hand Liu, Bu and Chen (2013) have studied DEM-CFD modelling of fluidized beds and implemented the Gidaspow (1994) drag model which uses the Wen-Yu (1966) equation for $\varepsilon \geq 0.8$ and the Ergun (1952) equation at smaller voidage. As such, it is a curious hybrid: the Ergun equation obtains from picturing flow through static packed beds as flow through tortuous capillaries. Clearly the capillary picture fails when the voidage exceeds the simple cubic packing value $\varepsilon_{sc} = 0.476$, and it comes as no surprise that the hybrid is discontinuous. Further, there is little reason to believe the static Ergun equation will apply in a dynamic context.

Computational advances - especially lattice Boltzmann methods - have encouraged a multi-scale modelling strategy, van der Hoef, van Sint Annaland, Deen and Kuipers (2008), and the determination of the drag force by fits to simulations of fluid flow through a random array of spheres. This program was initiated by Hill, Koch and Ladd (2001). The method has been extended to higher Reynolds number by Beetstra, Van der Hoef, and Kuipers (2007), who have also given a single global correlation in place of the rather complicated piecewise form of the Hill-Koch-Ladd drag force. More recently Tang, Peters, Kuipers, Kriebitzsch and van der Hoef (2014) have re-examined the lattice Boltzmann simulations, including mobility in the array which leads to an increase in the drag force. While this approach represents a considerable advance for fluidized bed modelling it is not a panacea: similar to settling based models, one is building many-body correlative effects into the single particle drag force, effects that may well be inappropriate for say a tumbling or stirred mill as there is no equivalent frame.

At vanishing Reynolds number, an alternative is provided by cell models of Happel (1958) and Kuwabara (1959). The advantage of the cell model for Eulerian – Lagrangian modelling is that the drag force only depends on the fluid in the immediate vicinity of the particle. That is to say one is protected against the inadvertent inclusion of collective effects into the drag force itself. The downside is that while numerical tabulations of the drag coefficient are available up to moderate $R_e$ for both, Juncu (2009), a convenient interpolating $C_D(R_e, \varepsilon)$ is not.

In this paper we will show that a simple, well-motivated, change to one of the Kuwabara boundary conditions enables the resulting "cell averaged model" to give an improved description of hindered settling at vanishing $R_e$ and small $\varepsilon$; this is done in Section 1. There we also show how collective effects enter the lattice Boltzmann drag models. The cell averaged model is extended to small but non-zero Reynolds number in Section 2 – the main motivation for this is to expose the relationship between the cell boundary conditions, cell size and Reynolds number. We then show in Section 3, using physical arguments about boundary layers versus boundary conditions, and supported by the Juncu simulations, that this model can be extended to $R_e \leq 10^3$

## 1. Stokes Cell Averaged Model

With the neglect of inertial terms, the Navier-Stokes equations for an incompressible Newtonian fluid having mass density $\rho$ and viscosity $\mu$,

$$\vec{\nabla} \cdot \vec{u} = 0, \quad \rho(\frac{\partial}{\partial t}\vec{u} + (\vec{u} \cdot \vec{\nabla})\vec{u}) = -\vec{\nabla}p - \mu\vec{\nabla} \times \vec{\omega}, \quad \vec{\omega} \equiv \vec{\nabla} \times \vec{u} \tag{1.1}$$

reduce to Stokes equations, for which a time-independent axially-symmetric solution is (Happel and Bremmer, 1965):

$$u_r = -(2c + \frac{2d}{r^3} - \frac{a}{r} + \frac{br^2}{5})\cos\theta \tag{1.2}$$

$$u_\theta = (2c - \frac{d}{r^3} - \frac{a}{2r} + \frac{2br^2}{5})\sin\theta . \tag{1.3}$$

Here the vorticity is

$$\omega_\phi = \sin\vartheta(\frac{a}{r^2} + br) . \tag{1.4}$$

The drag force on a spherical particle of radius $r_p = d_p/2$ is

$$\vec{F}_{drag} = -4\pi\mu a \hat{z} . \tag{1.5}$$

No-slip conditions at the surface of the particle moving with velocity $\vec{v}_p = v_p \hat{z}$ read:

$$2c + \frac{2d}{r_p^3} - \frac{a}{r_p} + \frac{br_p^2}{5} = -v_p = 2c - \frac{d}{r_p^3} - \frac{a}{2r_p} + \frac{2br_p^2}{5} . \tag{1.6}$$

We recall that for Stokes past an isolated stationary sphere $b = 0$; the constant $c$ is related to the fluid velocity at infinity, $\vec{u}(r \to \infty) = u_\infty \hat{z}$, and, eliminating $d$, to the constant $a$ appearing in the drag force:

$$-u_\infty = 2c = \frac{2a}{3r_p} - v_p \quad (\text{Stokes flow}) . \tag{1.7}$$

The canonical Stoke drag force then follows as

$$\vec{F}_{Stokes} = 6\pi\mu r_p \left( \vec{u}_\infty - \vec{v}_p \right) \quad \text{(Stokes flow)}. \tag{1.8}$$

In the cell model for concentrated systems, the particle is viewed as being surrounded by a concentric fluid envelope extending to radius $r_c$ that is related to the local voidage $\varepsilon$ and complimentary solid volume fraction $\phi$ by

$$1 - \varepsilon = \varphi = (r_p / r_c)^3. \tag{1.9}$$

The most widely used of these is the Happel (1958) cell model: the particle is moving at the centre at velocity $-\vec{u}_{rel} = \vec{v}_p - \vec{u}_f$, $\vec{u}_f$ being the interstitial fluid velocity, with free-surface conditions imposed at the outer boundary:

$$u_r\big|_{r_p} = 0, \quad \left[ r \frac{\partial}{\partial r}\left( \frac{u_\theta}{r} \right) + \frac{1}{r}\frac{\partial u_r}{\partial r} \right]_{r_p} = 0 \quad \text{(Happel)}. \tag{1.10}$$

That is to say

$$2c + \frac{2d}{r_p^3} - \frac{a}{r_p} + \frac{br_p^2}{5} = v_{rel} = 2c - \frac{d}{r_p^3} - \frac{a}{2r_p} + \frac{2br_p^2}{5} \quad \text{(Happel)} \tag{1.11}$$

$$2c + \frac{2d}{r_c^3} - \frac{a}{r_c} + \frac{br_c^2}{5} = \frac{2d}{r_c^4} + \frac{br_c}{5} = 0 \quad \text{(Happel)}. \tag{1.12}$$

The drag force in Happel's model follows as

$$\vec{F}_{drag} = 6\pi\mu r_p \left[ \frac{2}{3} \frac{3 + 2\phi^{5/3}}{2 - 3\phi^{1/3} + 3\phi^{5/3} - 2\phi^2} \right] \left( \vec{u}_f - \vec{v}_p \right) \quad \text{(Happel)}. \tag{1.13}$$

By contrast, Kuwabara's (1959) cell model assumes a stationary particle at the centre, with no-slip boundary conditions at $r_p$. The third boundary condition in the Kuwabara model arises from the vanishing of the vorticity at infinity in Stokes flow: one requires that the vorticity now vanish at the cell boundary:

$$\omega_\phi(r_c) = \sin\vartheta \left( \frac{a}{r_c^2} + br_c \right) = 0 \quad \text{(Kuwabara)}. \tag{1.14}$$

One other boundary condition is needed to find the drag force – Kuwabara, 1959 takes this as 'the radial fluid velocity at the cell boundary equals the radial component of the interstitial fluid velocity:

$$2c + \frac{2d}{r_c^3} - \frac{a}{r_c} + \frac{br_c^2}{5} = -u_f \quad \text{(Kuwabara)}. \tag{1.15}$$

Allowing the particle to also be in motion (1.6),(1.14) and (1.15) yield a drag force

$$\vec{F}_D = 6\pi\mu r_p \left[ \frac{1}{1+\phi - \tfrac{9}{5}\phi^{1/3} - \tfrac{1}{5}\phi^2} \right] (\vec{u}_f - \vec{v}_p) \quad \text{(Kuwabara)}. \tag{1.16}$$

While the zero-vorticity condition (1.14) has a clear physical motivation, Kuwabara's second boundary condition does not. There are two fluid velocities that have a well-defined meaning in multi-phase flow: the superficial velocity $\vec{U}$ and the associated interstitial velocity $\vec{u}_f$. We therefore suggest supplanting Kuwabara's second condition with[1]

$$\vec{U} = \varepsilon \vec{u}_f = \langle \vec{u} \rangle_{cell} \quad \text{(Cell Average)} \tag{1.17}$$

That is to say: the average fluid velocity over the cell volume *is* the superficial velocity. A simple calculation yields

$$\vec{U} = \left\{ -2c\left[1-\left(\frac{r_p}{r_c}\right)^3\right] + \frac{a}{r_c}\left[1-\left(\frac{r_p}{r_c}\right)^2\right] - \frac{br_c^2}{5}\left[1-\left(\frac{r_p}{r_c}\right)^5\right] \right\}\hat{z} \quad \text{(Cell Average)} \tag{1.18}$$

The drag force follows as:

$$\vec{F}_D = 6\pi\mu r_p \left[ \frac{\varepsilon}{1+\phi - \tfrac{9}{5}\phi^{1/3} - \tfrac{1}{5}\phi^2} \right] (\vec{u}_f - \vec{v}_p) \quad \text{(Cell Average)} \tag{1.19}$$

A convenient way to compare these models is in terms of the ratio of the terminal settling velocity in a suspension to that of isolated sphere:

$$v_r = \frac{v_t}{v_{t0}} \tag{1.20}$$

Since the bracketed factors in (1.13), (1.16) and (1.19) give the modification to the Stokes drag, it is trivial to read off that

$$v_r = \begin{cases} \dfrac{3}{2} \dfrac{2 - 3\phi^{1/3} + 3\phi^{5/3} - 2\phi^2}{3 + 2\phi^{5/3}} & \text{(Happel)} \\ 1 + \phi - \tfrac{9}{5}\phi^{1/3} - \tfrac{1}{5}\phi^2 & \text{(Kuwabara)} \\ \left(1 + \phi - \tfrac{9}{5}\phi^{1/3} - \tfrac{1}{5}\phi^2\right)/\varepsilon & \text{(Cell Average)} \end{cases} \tag{1.21}$$

---

[1] The cell-average model was previously described in Govender et al (2010), but without direct comparison to the Happel and Kuwabara models or to the settling data. Subsequently we learned that similar ideas had been discussed by Umnova et al (2000).

These expressions apply in an infinite system and for a quiescent fluid. In comparing to batch settling data, one should note that there the downward motion of the solids displaces fluid resulting in $\vec{u}_f = -\phi \vec{v}_p / \varepsilon$, so $\vec{u}_{rel} = -\vec{v}_p / \varepsilon$ and hence it is the slip velocity $\vec{U}_{slip} = \varepsilon \vec{u}_{rel} = \vec{U} - \varepsilon \vec{v}_p$ that corresponds to $-\vec{v}_p$. For the fluidized bed frame of quiescent particles and upward fluid flow the slip velocity coincides with the superficial velocity; there at low Reynolds number, Richardson and Zaki (1954) give

$$v_r = \varepsilon^{4.65} \quad (\text{Richardson-Zaki}). \tag{1.22}$$

The cell model results are compared to (1.22) in Figure 1. There we also give $v_r$ according to the creeping flow truncation of the Ergun equation:

$$\vec{F}_D = 6\pi\mu r_p \left[ \frac{25}{3} \frac{1-\varepsilon}{\varepsilon^2} \right] (\vec{u}_\infty - \vec{v}_p) \quad (\text{Truncated Ergun}). \tag{1.23}$$

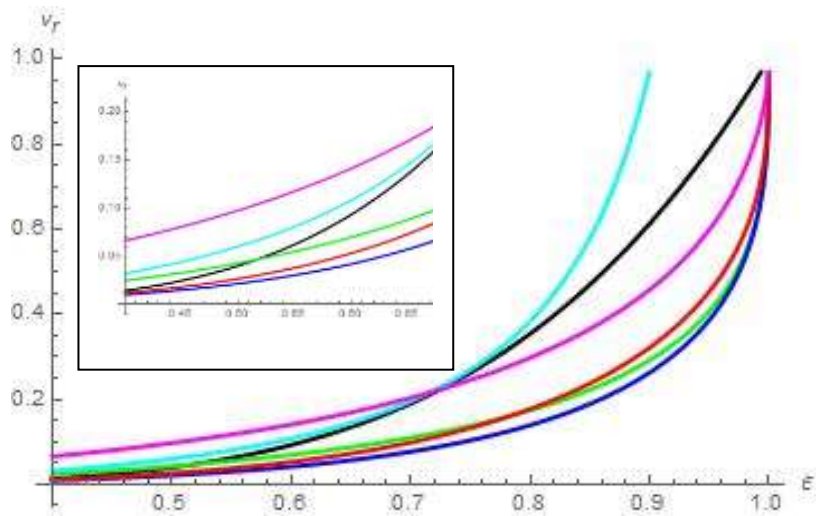

Figure 1: Relative settling velocities versus voidage at vanishing Reynolds number for Richardson-Zaki (black), cell averaged (green), Kuwabara (blue), Happel (red), truncated Ergun (cyan), and truncated Beetstra (magenta) models.

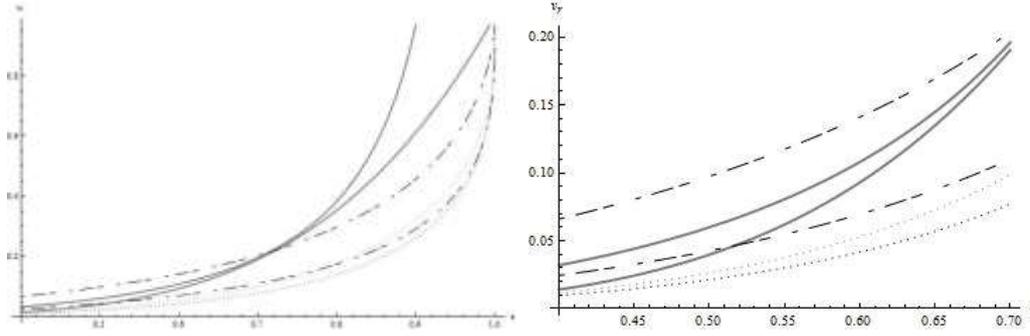

Figure 1: Relative settling velocities versus voidage at vanishing Reynolds number for Richardson-Zaki (thick grey), cell averaged (short dashed), Kuwabara (medium dashed), Happel (long dashed), truncated Ergun (dotted grey), and truncated Beetstra (dotted) models.

The three cell models are grossly similar in that they badly miss the Richardson-Zaki curve at large voidage. As pointed out by Batchelor (1972) this is inevitable in that cell models do not encompass the statistical fluctuations that are responsible for the observed $v_r - 1 \propto \phi$ at small solid volume fraction. A clear distinction arises, however, at small voidage. The cell averaged model is the only one that comes close to the canonical Ergun equation at the nominal packed bed $\varepsilon = 0.4$ but, unlike the latter, does not involve a fit parameter like tortuosity. Moreover, the simple cubic $\varepsilon_{sc} = 0.476$ constitutes a critical point: at smaller voidage occupied-volume effects are important for hindered settling, whereas at larger $\varepsilon$ collective fluid effects are crucial. Just above $\varepsilon_{sc}$, however, the close proximity of the particles assures that the flow around each will be statistically similar such that the single-particle cell model drag should account for hindered settling. Thus it is decisive that the cell averaged model is the only one that approximately describes the data encapsulated by (1.22) in the interval $0.48 \leq \varepsilon \leq 0.54$.

In Figure 1 we have also displayed the relative settling velocity obtained from the creeping flow truncation of the Beetstra, Van der Hoef, and Kuipers (2007) drag model:

$$\vec{F}_D = 6\pi r_p \mu \left[ \frac{10\phi}{\varepsilon} + \varepsilon^3 \left(1 + 1.5\sqrt{\phi}\right) \right] \quad \text{(Truncated Beetstra)} \qquad (1.24)$$

What is first notable is that at low voidage the Beetstra drag gives the highest settling velocity of all the models. Moreover, the Beetstra model does intersect the Richardson-Zaki curve at $\varepsilon \simeq 0.72$ - this can be understood as reflecting that the lattice Boltzmann simulations of fluid flow through an random array of spheres do pick up collective fluid effects which are then included in the "single particle" drag force.

## 2. Oseen Cell Averaged Model

There are a limited number of tools available (Veysey and Goldenfeld, 2007) to extend the cell model to small Reynolds number

$$R_e \equiv \frac{\rho d_p |\vec{u}_f - \vec{v}_p|}{\mu}. \tag{2.1}$$

The one best suited is the Oseen approximation: the Navier-Stokes equations are linearized in

$$\delta\vec{u} = \vec{u} - \vec{u}_{rel}; \tag{2.2}$$

Oseen's equations are

$$\vec{\nabla}\cdot\delta\vec{u} = 0, \quad \rho(\frac{\partial}{\partial t}\delta\vec{u} + (\vec{u}_{rel}\cdot\vec{\nabla})\delta\vec{u}) = -\vec{\nabla}p - \mu\vec{\nabla}\times\delta\vec{\omega}, \quad \delta\vec{\omega} \equiv \vec{\nabla}\times\delta\vec{u}. \tag{2.3}$$

The standard strategy for solving (2.3) is to take $\vec{u}_{rel} = u_{rel}\hat{z}$, and write $\delta\vec{u}$ as

$$\delta\vec{u} = \vec{\nabla}\Phi + \psi\hat{z} - \frac{\vec{\nabla}}{2k}\psi. \tag{2.4}$$

Here $k = \rho u_{rel}/2\mu$. The irrotational potential $\Phi$ is harmonic and determines the pressure:

$$\vec{\nabla}^2\Phi = 0 \tag{2.5}$$

$$p = -\rho u_{rel}\frac{\partial\Phi}{\partial z}. \tag{2.6}$$

The potential $\psi$ can be re-expressed in terms of a potential satisfying Helmholtz' equation:

$$\psi = e^{kz}\Psi, \quad (\vec{\nabla}^2 - k^2)\Psi = 0. \tag{2.7}$$

We consider a particular solution of (2.7) in term of spherical modified Bessel functions,

$$\Psi = AK_0(kr) + BI_0(kr) \tag{2.8}$$

This gives rise to the vorticity:

$$\omega_\varphi = k\sin\theta e^{kr\cos\theta}[AK_1(kr) - BI_1(kr)]. \tag{2.9}$$

Applying the outer boundary condition,

$$B = AK_1(kr_c)/I_1(kr_c). \tag{2.10}$$

The potential $\Phi$ is given as an infinite series in Legendre polynomials:

$$\Phi = \sum_{n=0}^{\infty}\left(\frac{C_n}{r^{n+1}} + D_n r^n\right)P_n(\cos\theta). \tag{2.11}$$

Indeed, an infinite number of terms are required to enforce the no-slip boundary condition $\delta\vec{u}(r_p) = -\vec{u}_{rel}$. Fortunately, our interest lies with the drag force which, owing to the no-slip condition, comes solely from the vorticity as in the Stokes case:

$$\vec{F}_D = \int(-pd\vec{a} + \mu\delta\vec{\omega}\times d\vec{a}) = -4\pi\mu\frac{A}{k}\hat{z}. \quad (2.12)$$

Moreover, the cell averaged condition, which here becomes $\langle\delta\vec{u}\rangle_{cell} = 0$, leads to

$$\varepsilon D_1 + \frac{3}{2r_c^3}\int_{r_p}^{r_c}\{I_1(kr)[AK_1(kr) - BI_1(kr)] + I_0(kr)[AK_0(kr) + BI_0(kr)]\}dr = 0. \quad (2.13)$$

We thus need only the projected no-slip condition:

$$\begin{aligned}-u_{rel} &= \tfrac{1}{2}\int_0^\pi \hat{z}\cdot\delta\vec{u}(r_p)\sin\theta d\theta \\ &= D_1 + \tfrac{1}{2}\{I_1(kr_p)[AK_1(kr_p) - BI_1(kr_p)] + I_0(kr_p)[AK_0(kr_p) + BI_0(kr_p)]\}\end{aligned} \quad (2.14)$$

Taken together (2.10) (2.12), (2.13) and (2.14) suffice to determine the Oseen cell model drag force. Albeit a general expression can be given in terms of spherical modified Bessel functions, it is rather opaque. Since Oseen's equations are in any case only an approximation, it is better to look at an expansion in the Reynolds number. There are two subcases; in the concentrated case $kr_c = R_e/4\phi^{1/3} \ll 1$

$$\vec{F}_D = \frac{6\pi\mu r_p(\vec{u}_f - \vec{v}_p)\varepsilon}{\left\{1 + \phi - \tfrac{9}{5}\phi^{1/3} - \tfrac{1}{5}\phi^2 + \dfrac{R_e^2}{80}\left[2 - \tfrac{1}{4}\phi - \tfrac{1}{7}\phi^2 + \tfrac{3}{5}\phi^{4/3} - \tfrac{3}{2}\phi^{1/3} - \tfrac{99}{140}\phi^{-1/3}\right]\right\}}, \quad (2.15)$$

$$R_e \ll 4\phi^{1/3}$$

The ratio of (2.15) and Stokes cell averaged models is displayed in Figure 2. Since the effect is tiny, and the Reynolds number quite limited, the Oseen cell averaged model is mainly of academic interest.

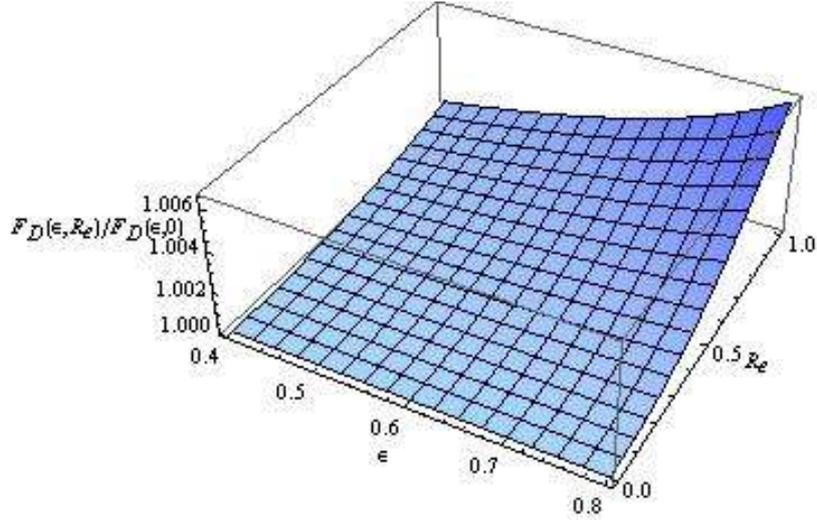

Figure 2: Ratio of the Oseen cell averaged model drag force (2.15) to the Stokes cell averaged model drag force (1.19), for $0.4 \leq \varepsilon \leq 0.8$ and $R_e \leq 1$.

In the dilute limit $kr_c = R_e/4\phi^{1/3} \gg 1$ (2.10) becomes

$$B \simeq Ae^{-2kr_c} \simeq 0, \quad kr_c = R_e/4\phi^{1/3} \gg 1. \tag{2.16}$$

That is to say, the outer boundary condition of the cell averaged model becomes irrelevant since the vorticity is exponentially suppressed outside a radius of $k^{-1} = 2d_p/R_e$ from the particle's centre that is small compared to $r_c$. One then obtains a correction to the canonical Oseen drag force :

$$\vec{F}_D = \frac{6\pi\mu r_p (\vec{u}_f - \vec{v}_p)}{\left\{1 - \dfrac{3R_e}{16} - \dfrac{9\phi^{2/3}}{R_e}\right\}}, \quad \phi^{1/3} \ll R_e \ll 1. \tag{2.17}$$

### 3. Inertial Cell Model

For all its limitations, what the Oseen approximation does capture is that viscous effects have a finite range. At larger Reynolds number one expects the onset of boundary layer formation: the outer boundary condition of the cell averaged model will become irrelevant. This observation is particularly suggestive since the drag force can be expressed in term of the dimensionless drag coefficient

$$\vec{F}_D \equiv \frac{\rho}{8} d_p^2 C_D(R_e) |\vec{u}_f - \vec{v}_p| (\vec{u}_f - \vec{v}_p), \tag{3.1}$$

and one notes that in the relevant range of Reynolds number the isolated sphere drag coefficient is well approximated by (Bird et al 1960)

$$C_D = \left(\sqrt{\frac{24}{R_e}} + 0.54\right)^2, R_e \leq 6\cdot 10^3, \varepsilon = 1. \qquad (3.2)$$

In the Stokes limit the cell averaged model gives

$$C_D \simeq \frac{24}{R_e}\left[\frac{\varepsilon}{1+\phi-\tfrac{9}{5}\phi^{1/3}-\tfrac{1}{5}\phi^2}\right], \quad R_e \ll 1 \,(\text{Stokes cell average}) \qquad (3.3)$$

Owing to boundary layer effects, we hypothesize that $C_D$ has a similar functional form to (3.2) while matching to the cell averaged model as $R_e \to 0$:

$$C_D(R_e,\varepsilon) = \left(\sqrt{\frac{24}{R_e}\frac{\varepsilon}{(1+\phi-\tfrac{9}{5}\phi^{1/3}-\tfrac{1}{5}\phi^2)}} + 0.54\right)^2 \quad (\text{Inertial Cell Model}). \qquad (3.4)$$

The crucial question is how to test this hypothesis.

Juncu (2009) has numerically calculated the drag coefficient at various voidages, and up to moderate Reynolds number, in both the Happel and Kuwabara cell models. It must be noted that what Juncu refers to as the superficial fluid velocity is, according to his non-dimensional boundary conditions, in fact the interstitial fluid velocity. For the Kuwabara model the inertial hypothesis reads:

$$C_D(R_e,\varepsilon) = \left(\sqrt{\frac{24}{R_e}\frac{1}{(1+\phi-\tfrac{9}{5}\phi^{1/3}-\tfrac{1}{5}\phi^2)}} + 0.54\right)^2 \quad (\text{Inertial Kuwabara Model}) \qquad (3.5)$$

A comparison of (3.5) with Juncu's numerical data is shown in figure 3. Since no fit was involved, while (3.5) maps into (3.4) via $R_e \to R_e/\varepsilon$, we can take this as validation of the Inertial Cell Model.

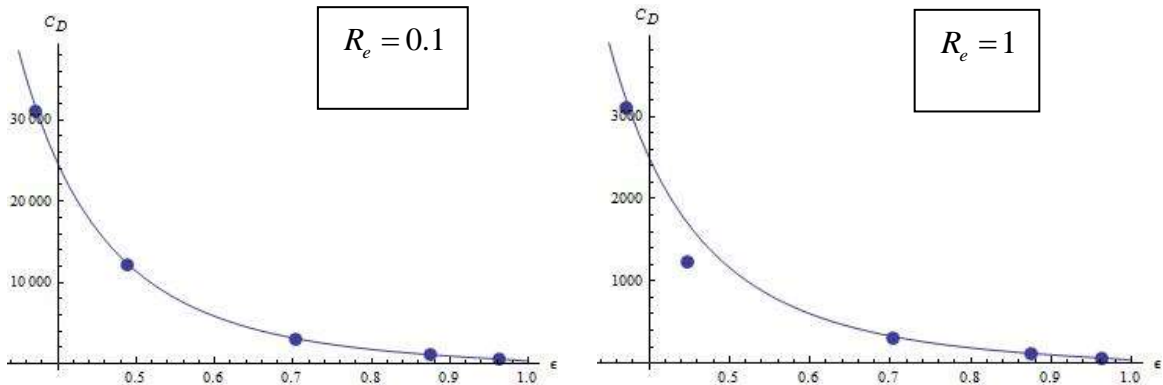

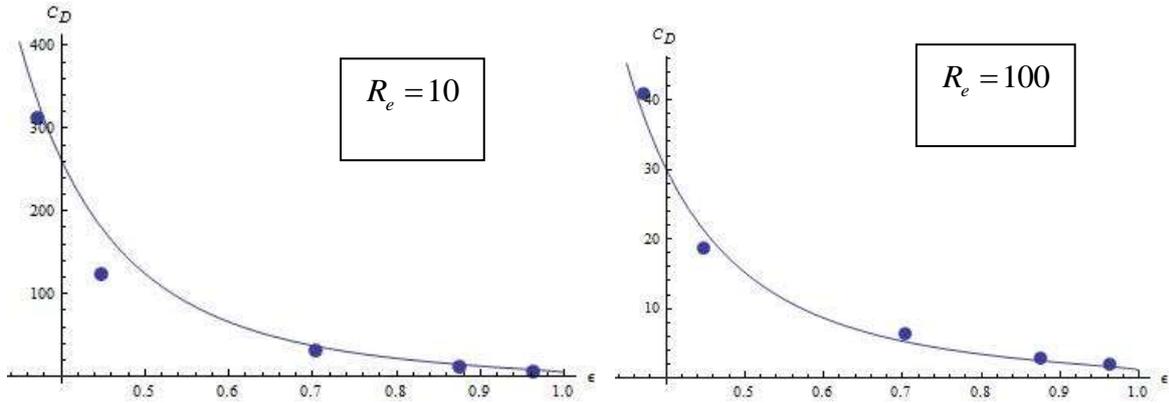

Figure 3: Comparison of the inertial Kuwabara cell model drag coefficient (3.5) with the numerical data of Juncu (2009).

In Figure 4 we exhibit the ratio of the Inertial Cell Model drag coefficient to that of an isolated sphere. What is evident there is that the use of a single sphere drag is only justified at large voidage and high Reynolds number.

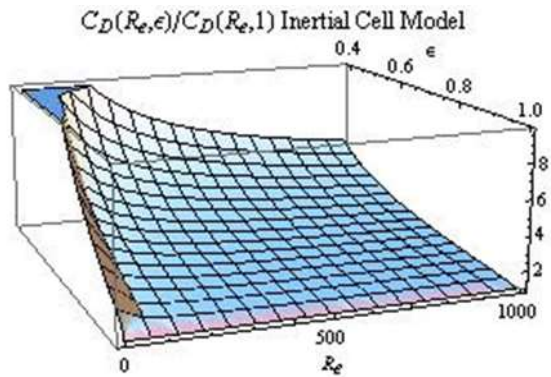

Figure 4: Ratio of the Inertial Cell Model drag coefficient to that of a single sphere.

The drag coefficient (3.4) corresponds to a drag force

$$\vec{F}_D = 6\pi r_p \mu (\vec{u}_f - \vec{v}_p) \left[ \sqrt{f(\varepsilon)} + 0.11\sqrt{R_e} \right]^2 \quad \text{(Inertial Cell Model)}$$

$$f(\varepsilon) \equiv \frac{\varepsilon}{1 + \phi - \frac{9}{5}\phi^{1/3} - \frac{1}{5}\phi^2}$$

(3.6)

This drag force, in units of the Stokes drag, is shown in figure 5. Therein, for comparison, we also exhibit the drag force obtained by Beetstra et el (2007) based upon fits to lattice Boltzmann simulations:

$$\vec{F}_D = 6\pi r_p \mu (\vec{u}_f - \vec{v}_p) \left[ \frac{10\phi}{\varepsilon} + \varepsilon^3 \left(1 + 1.5\sqrt{\phi}\right) + \frac{0.413 R_e}{24} \left\{ \frac{\varepsilon^{-1} + 3\varepsilon\phi + 8.4(\varepsilon R_e)^{-0.343}}{1 + 10^{3\phi}(\varepsilon R_e)^{-(1+4\phi)/2}} \right\} \right] \quad (3.7)$$

(Beetstra et al, 2007)

Although the two models give similar results, the Inertial Cell Model gives a larger drag force – and hence smaller terminal settling velocity - at low voidage.

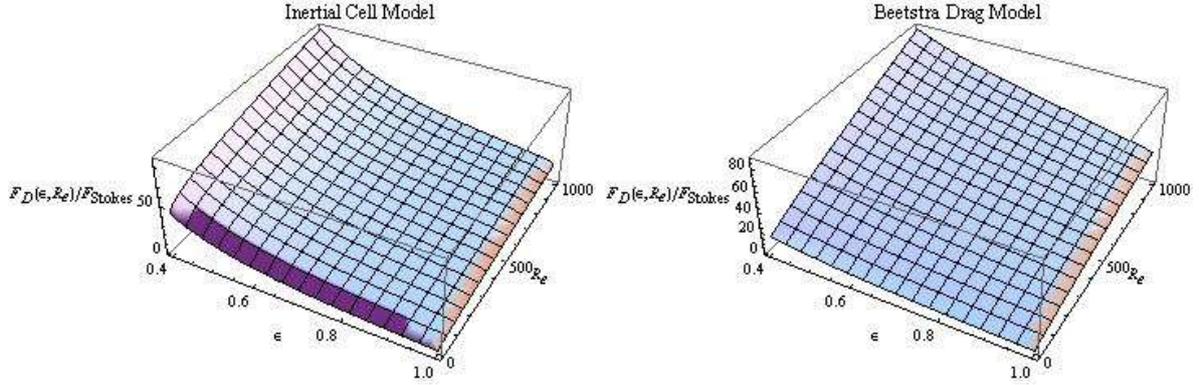

Figure 5: Drag force in units of the Stokes drag for the Inertial Cell Model and the model of Beetstra et al.

A nice feature of the Inertial Cell Model is that it allows one analytic expressions for e.g. the terminal settling velocity:

$$0 = \vec{F}_D + \frac{d_p^3}{6} \Delta \rho \vec{g} \,. \quad (3.8)$$

Using (3.6), in terms of $R_t = \rho d_p v_t / \mu$ and the Archimedes number $A_r = d_p^3 \rho \Delta \rho g / \mu^2$ we have

$$\frac{A_r}{18} = R_t \left[ \sqrt{f(\varepsilon)} + 0.11 \sqrt{R_t} \right]^2 \quad (3.9)$$

The solution in our model is:

$$v_t = \frac{\mu}{0.0484 \rho d_p} \left\{ -\sqrt{f(\varepsilon)} + \sqrt{f(\varepsilon) + 0.44 \sqrt{\frac{A_r}{18}}} \right\}^2, \quad (3.10)$$

while for an isolated sphere

$$v_{t0} = \frac{\mu}{0.0484 \rho d_p} \left\{ -1 + \sqrt{1 + 0.44 \sqrt{\frac{A_r}{18}}} \right\}^2. \quad (3.11)$$

Eliminating the Archimedes number gives ratio of the terminal settling velocity in a suspension to that of isolated sphere:

$$v_r = \frac{1}{0.0484 R_{t0}} \left\{ -\sqrt{f(\varepsilon)} + \sqrt{f(\varepsilon) + 0.44\sqrt{R_{t0}} + 0.0484 R_{t0}} \right\}^2. \qquad (3.12)$$

We have chosen to parameterize in terms of the isolated terminal Reynolds number since this appears in the Richardson-Zaki model:

$$v_r = \varepsilon^n, \; n = \begin{cases} 4.65, \; R_{t0} < 0.2 \\ 4.4 R_{t0}^{-0.03}, \; 0.2 < R_{t0} < 1 \\ 4.4 R_{t0}^{-0.1}, \; 1 < R_{t0} < 500 \\ 2.4, \; 500 < R_{t0} \end{cases} \text{(Richardson-Zaki)} \qquad (3.13)$$

In figure 6 we show the relative settling velocities (3.12) and (3.13). One observes that even at comparatively small $\varepsilon$, but moderate $R_e$, the Inertial Cell Model gives a higher $v_r$. As in the opposite case of small $R_e$ and moderate $\varepsilon$ discussed earlier, this arises because the model does not see collective fluid effects.

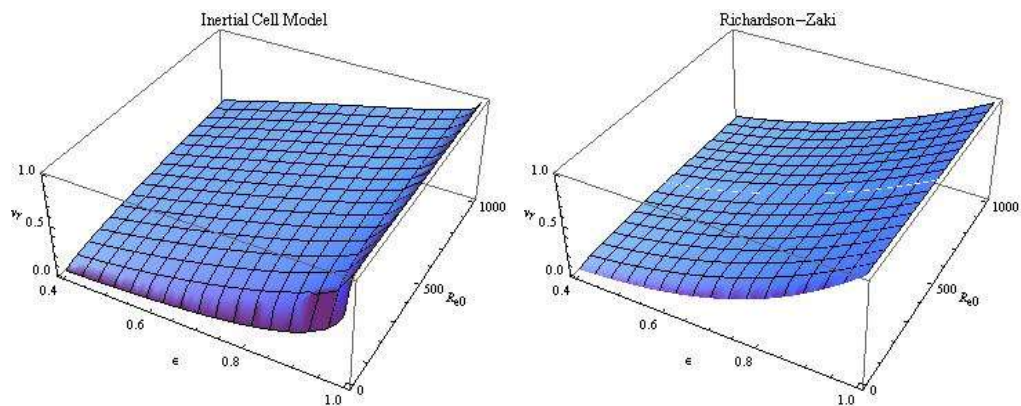

Figure 6: Relative terminal settling velocity in the Inertial Cell and Richardson-Zaki models.

Finally, it is illuminating to compare the Inertial Cell Model (3.6) with the full Ergun (1952) equation as used in Gidaspow (1994):

$$\vec{F}_D = 6\pi\mu r_p \left[ \frac{25}{3} \frac{1-\varepsilon}{\varepsilon^2} + \frac{7}{72} \frac{R_e}{\varepsilon} \right] (\vec{u}_\infty - \vec{v}_p) \quad (\text{Ergun}). \qquad (3.14)$$

The ratio of the forces (3.6)/(3.14) is displayed in Figure7. The differences aptly illustrate the perils of employing a static model in a dynamic context.

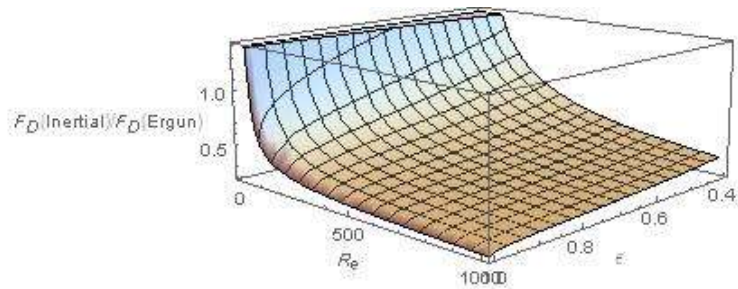

Figure 7: Ratio of the drag force in the Inertial Cell Model (3.6) to that given by the Ergun (1952) equation (3.14).

## 4. Conclusions

In this paper we have taken a cell model approach to the drag force in solid- fluid multiphase flow, with the particular aim of isolating local from collective fluid effects. In the Stokes limit we demonstrated that the cell averaged model is singled out as the only one to account for hindered settling near the simple cubic packing voidage. We then extended the cell averaged model to small Reynolds number in the Oseen approximation; this allowed us to examine the interplay of Reynolds number and voidage in the model's zero-vorticity boundary condition. Keying from the latter we made the inertial hypothesis to extend the model to moderate Reynolds number, and checked this against available numerical results.

Our main result is the Inertial Cell model drag coefficient(3.4)/drag force(3.6) which is free of extraneous collective effects, and so is as suitable for Eulerian – Lagrangian modelling of cyclones and mills as it is for fluidized bed. It provides also a dynamic replacement for the continued (mis)use of the static Ergun (1952) equation in Eulerian – Eulerian modelling.

## Nomenclature

a,b,c,d   Coefficients in Stokes cell model

A,B,C,D   Coefficients in Oseen cell model

$A_r$   Archimedes number

$C_D$   Drag coefficient

$d_p$   particle diameter

$F_D$   Drag force

k   Radial wavenumber in Oseel cell model

$r_c$  Cell radius

$r_p$  Particle radius

$R_e$  Reynolds number

$R_t$  Terminal settling Reynolds number

$u$  Fluid velocity

$u_f$  Interstitial fluid velocity

$U$  Superficial fluid velocity

$U_{slip}$  Slip velocity

$v_p$  particle velocity

$v_t$  Terminal settling velocity

*Greek*

$\varepsilon$  Voidage

$\phi$  Solid volume fraction

$\mu$  Fluid viscosity

$\rho$  Fluid density

## Acknowledgements

This work was supported by Anglo-American Platinum. The work of G. B. Tupper was partially supported by a grant from the National Research Foundation.